\documentclass[%
mathleft,%
final,
]{an}
\usepackage{graphicx}
\usepackage[varg]{txfonts}
\usepackage{natbib}
\bibpunct{(}{)}{;}{a}{}{,}
\usepackage{url}
\newcommand{\kms}{km\,s$^{-1}$}

\newcommand{\COBOLD}{{\sf CO$^5$BOLD}}
\newcommand{\cobold}{\COBOLD}
\newcommand{\linfor}{{\sf Linfor3D}}
\newcommand{\nlte}{{\sf NLTE3D}}
\def\aj{AJ}%
\def\apj{ApJ}%
\def\apjs{ApJS}%
\def\aap{A\&A}%
\def\aapr{A\&A~Rev.}%
\def\aaps{A\&AS}%
\def\mnras{MNRAS}%
\def\solphys{Sol.~Phys.}%
\def\nat{Nature}%
\def\gca{Geochim.~Cosmochim.~Acta}%
\def\memsai{Mem.~Soc.~Astron.~Italiana}%
\def\procspie{Proc.~SPIE}%

\def\teff{$T_{\rm eff}$}
\def\kms{$\mathrm{km\, s^{-1}}$}

\def\vsini {$v\sin\,i$}
\def\vrad {$\mathrm{v\,_{\mathrm rad}}$}
\newcommand{\logg}{\ensuremath{\log g}}
\begin{document}

\Pagespan{789}{}
\Yearpublication{2015}%
\Yearsubmission{2015}%
\Month{11}%
\Volume{999}%
\Issue{88}%
\DOI{This.is/not.aDOI}%

\title{Chemical composition of a sample of bright solar-metallicity stars\,\thanks{Data obtained at Observatoire de Haute
Provence, with the SOPHIE spectrograph.}}

\author{E. Caffau\inst{1},\fnmsep\thanks{Corresponding author:
  \email{Elisabetta.Caffau@obspm.fr}\newline}
A. Mott\inst{2},
M. Steffen\inst{2,1},
P. Bonifacio\inst{1},
K. G. Strassmeier\inst{2},
A. Gallagher\inst{1},
R. Faraggiana\inst{3},
L. Sbordone\inst{4,5}
}
\titlerunning{Chemical composition of a sample of bright solar metallicity stars}
\authorrunning{E. Caffau et al.}
\institute{
GEPI, Observatoire de Paris, PSL Resarch University, CNRS,
Universit\'e Paris Diderot, Sorbonne Paris Cit\'e, Place Jules Janssen, 92195
Meudon, France
\and
Leibniz-Institut f{\"u}r Astrophysik Potsdam, An der Sternwarte 16,
D-14482 Potsdam, Germany
\and
Stradella Verde 82 - 34079 Staranzano (GO), Italy
\and
Millennium Institute of Astrophysics, Av. Vicu\~na Mackenna 4860, 782-0436 Macul, Santiago, Chile
\and
Pontificia Universidad Cat\'olica de Chile, Av. Vicu\~na Mackenna 4860, 782-0436
Macul, Santiago, Chile
}

\received{7 August 2015}
\accepted{xx xxx xxxx}
\publonline{later}

\keywords{Galaxy: abundances - stars: abundance -- stars: activity -- stars: atmospheres}

\abstract{%
We present a detailed analysis of seven young stars observed with the
spectrograph SOPHIE at the Observatoire de Haute-Provence for which
the chemical composition was incomplete or absent in the literature.
For five stars, we derived the stellar parameters and chemical
compositions using our automatic pipeline optimized for F, G, and K
stars, while for the other two stars with high rotational velocity, we
derived the stellar parameters by using other information (parallax),
and performed a line-by-line analysis. Chromospheric emission-line fluxes from \ion{Ca}{ii} are obtained for all targets.
The stellar parameters we derive are generally in good agreement with
what is available in the literature.
We provide a chemical analysis of two of the stars for the first time.
The star HIP\,80124 shows a strong Li feature at 670.8\,nm implying a
high lithium abundance.  Its chemical pattern is not consistent with
it being a solar sibling, as has been suggested.}

\maketitle

\section{Introduction}

Active stars are late-type stars which show emission components of
resonance lines in their spectra, which originate in chromospheric
layers of the stellar atmosphere. In the solar atmosphere, line
emissions appear in regions where magnetic fields are strong, so the
presence of these fields must be linked to the solar activity.
Several properties of these stars have been analysed
\citep{strassmeier09,mamajek08,pizzolato03}. For example, correlation
of activity with magnetic fields (e.g. \citealt{montesinos87}); with
rotational periods (e.g. \citealt{karak14,takeda10,suarez15,montes01}),
(but we have to warn that \vsini\ does not strictly mean rotation period); inverse
correlation with age due to decrease of stellar rotation with age
(e.g. \citealt{lopez10}); binarity (e.g \citealt{eker08}); and
relation with Li abundance (e.g. \citealt{strassmeier12,takeda10,israelian09,delgadomena14,delgadomena15}).
We found no complete abundance analyses in the literature for some of the bright
members of this class of stars. Accurate metallicity determination is
important to derive stellar age, when this is based on isochrones, so
that one can use the isochrones at the right metallicity. Their
chemical composition is also important to put these stars in the
correct place in the chemical evolution in the Milky Way and in the
investigation of atmospheric phenomena, such as diffusion.

The sample of active stars for which a detailed chemical analysis
is lacking or missing is large, but our choice of the sample we analysed
was constrained in brightness and
coordinates by the telescope size and assigned
telescope time we had.
Among the interesting candidates, we selected the ones with the constraint
to be observable at the beginning of March, in the last part of the night,
and being bright enough to give a quality spectrum
sufficient for a chemical analysis with the spectrograph SOPHIE on a 2\,m class telescope.
We present the chemical analysis of seven bright stars observed at
Observatoire de Haute-Provence (OHP): HIP\,67344, HIP\,74858,
HIP\,75132, HIP\,81284, HIP\,90864, HIP\,80124, and $\eta$\,UMi.

\section{Observations}

The spectra were obtained between the 1st and 5th March 2014 at
Observatoire de Haute-Provence, using the SOPHIE spectrograph
\citep{Bouchy,Perru08,Perru11} with the High Efficiency (HE) fibers,
that provide a resolving power in the range between 43\,000 and 37\,000,
depending on the position along the
order\footnote{http://www.obs-hp.fr/guide/sophie/thorium\_psf.html}.
The spectra cover the wavelength range from 387.2\,nm to 694.3\,nm.
We have a single spectrum for each program star except for HIP\,80124
for which we have two observations.  The time of the observations and
exposure times are provided in Table\,\ref{obslog}.  In the table we provide also the
radial velocity (RV) derived by the Geneva pipeline, along with its root mean square
error. We used this RV to shift the spectra before analysing
them with our automatic pipeline. We need a precision in RV of the order of the \kms\
for the chemical analysis and the pipeline of SOPHIE is largely adapted for this.
The pipeline derives also the projected rotational velocity (\vsini)
that we provide although we did not use it.
The method for deriving \vsini\ is based on the width
of the cross-correlation function and is described in detail in
\citet[][Appendix B]{Boisse}, the expected accuracy for the SOPHIE
High Resolution (HR) fibres is 1\,\kms. No estimate is provided for the
HE mode that we use. Since the resolving power in HE mode is
roughly one half of that in HR mode we think that it is justified to assume an error
of 2\,\kms\ on the \vsini\ estimates.
We also stress that the above method has been shown to
provide robust results for \vsini\ up to  20\,\kms,
but it is not well calibrated for fast rotators, in which cases the errors
can be much larger.
For HIP\,75132 (HD\,136655) the pipeline failed to determine \vsini\
and for HIP\,79022 (eta\,UMi) it gave an unrealistically high value.

\begin{table*}
\caption{Observations log.}
\label{obslog}
\setlength{\tabcolsep}{5pt}
\begin{tabular}{llcccrrrr}\hline
\hline
star                     &  V        &   UT &  UT       &  MJD         &$t_{\rm exp}$ &  RV          &  \vsini  & S/N \\
HIP (HD)                 &  &  date      &  h:m:s.sss   &  days        &  s           & km/s         & km/s & \@ order 15 \\
\hline
Ceres                    &      & 2014-03-02 & 02:12:37.657 & 56718.092106 &  600 & $  7.1627\pm 0.0015$ &   3.7  &  \\
HIP\,67344 (HD\,120205)  & 8.34 & 2014-03-05 & 02:11:44.823 & 56721.611164 & 2700 & $-29.2866\pm 0.0006$ &   4.6  & 152 \\
HIP\,74858 (HD\,136137)  & 6.65 & 2014-03-02 & 04:08:49.329 & 56718.678534 &  600 & $ +2.2891\pm 0.0007$ &   6.2  & 145 \\
HIP\,75132 (HD\,136655)  & 9.02 & 2014-03-05 & 03:00:24.486 & 56721.648568 & 3600 & $-32.3446\pm 0.0006$ &        & 150 \\
HIP\,81284 (HD\,150202)  & 7.97 & 2014-03-06 & 03:05:39.845 & 56722.650633 & 3600 & $-26.9423\pm 0.0005$ &   7.8  & 195 \\
HIP\,90864 (HD\,171067)  & 7.19 & 2014-03-04 & 03:46:59.156 & 56720.678779 & 4000 & $-46.2326\pm 0.0013$ &   3.8  &  81 \\
HIP\,80124 (HD\,147443)  & 8.71 & 2014-03-02 & 03:52:49.614 & 56718.651949 & 1800 & $ -6.5427\pm 0.0077$ &  38    & 110 \\
HIP\,80124 (HD\,147443)  & 8.71 & 2014-03-04 & 03:42:41.831 & 56720.634653 & 3600 & $ +4.1704\pm 0.0134$ &  34    &  74 \\
HIP\,79822 ($\eta$\,UMi) & 4.95 & 2014-03-06 & 04:08:19.898 & 56722.172454 &  300 & $-20.0841\pm 0.0085$ &  85$^a$& 258 \\
\hline
\end{tabular}
\\
{$^a$\,this value comes from our analysis.}
\end{table*}

\section{Chemical analysis}

\begin{figure*}
\includegraphics*[width=160mm]{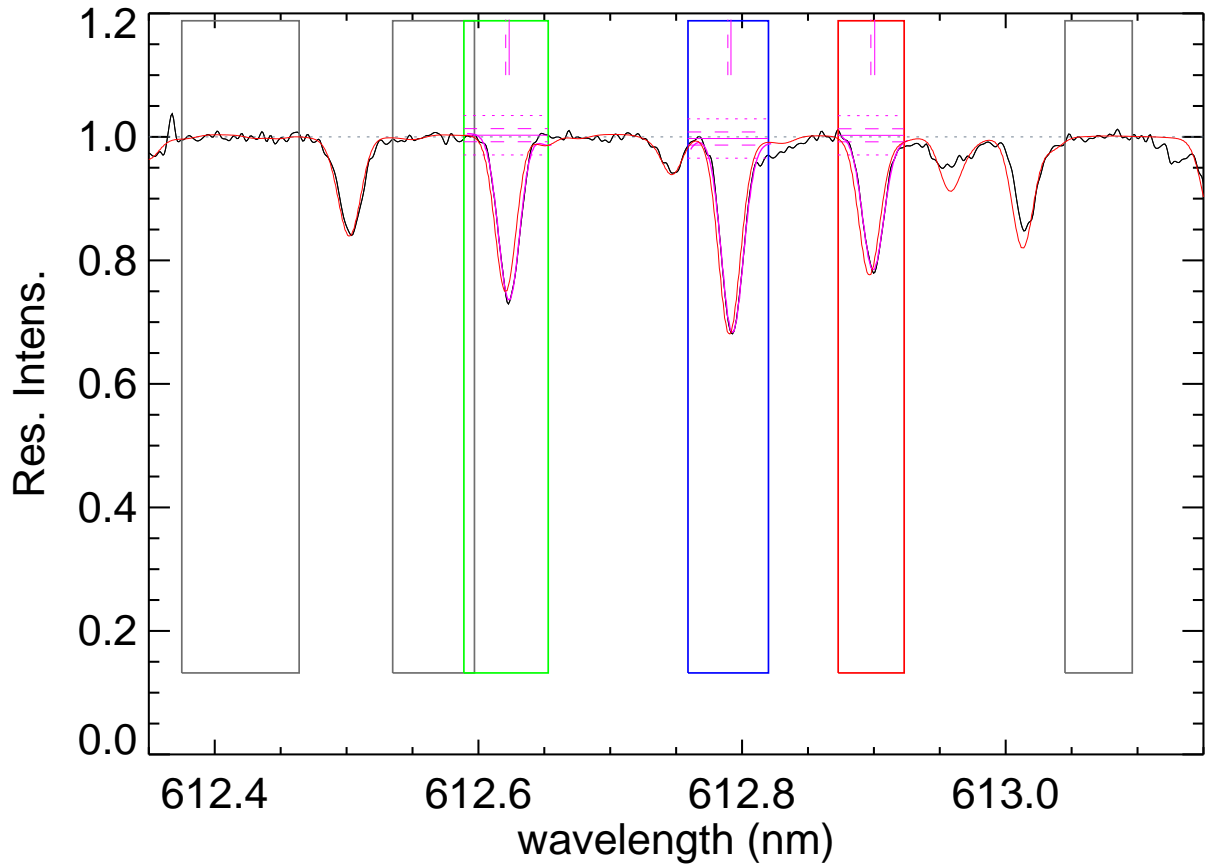}
\caption{The result of MyGIsFOS is here shown for HD\,120205.
Solid black line represent the observed spectrum, solid red the
synthetic spectrum with the final parameters, solid magenta is the best fit
profile for selected features.
Different colours of the rectangular boxes indicates different features:
in grey the ranges to pseudo-normalise the spectra;
in green, at 612.6\,nm, a \ion{Ti}{i} line;
in blue, at 612.8\,nm, a \ion{Fe}{i} line;
in red, at 612.9\,nm, a \ion{Ni}{i} line.
The magenta horizontal lines represents: the determined continuum (solid line)
and 1$\sigma$ of the S/N ratio (dashed line);
the magenta vertical solid lines represent the laboratory
wavelength, the dashed lines the shift in the line profile fitting.}
\label{mygisfos}
\end{figure*}

A sub-sample of our spectra have been analysed with the pipeline
MyGIsFOS \citep{sbordone14}.  This code simulates the traditional
analysis in an automatic way; it derives atmospheric parameters and
detailed stellar abundances.  The analysis done by MyGIsFOS proceeds
as follows:

\begin{itemize}
\item
The effective temperature, \teff, is derived from a set of isolated
\ion{Fe}{i} lines. A null slope is imposed to the linear fit of the
lower energy of the \ion{Fe}{i} lines versus the abundance.
\item
The micro-turbulence, $\xi_{\rm turb}$, is derived by imposing a null
slope to the equivalent width (EW) of isolated \ion{Fe}{i} lines as a
function of the abundance derived from these lines.
\item
log\,g is determined from \ion{Fe}{i} and \ion{Fe}{ii} ionisation
equilibrium.
\item
Fe abundance is determined from \ion{Fe}{i} lines.
\item
The enhancement of $\alpha$-elements is determined by measuring lines
of various $\alpha$ elements, and using their average abundance ratio
[X/Fe], as {$\left[{\rm X/Fe}\right]=\log\left[{N}\left(
      {\rm X}\right)/{N}\left({\rm Fe}\right)\right]_{\star}- \log\left[
      {N}\left({\rm X}\right)/{N}\left({\rm Fe}\right)\right]_\odot$} as an
estimate of [$\alpha$/Fe].
\item
All [X/H] for all X element represented in the spectrum are derived
from lines of \ion{X}{i} or \ion{X}{ii}.
\end{itemize}
An example of the analysis done by MyGIsFOS is shown in Fig.\,\ref{mygisfos}.

Five stars (HIP\,67344, HIP\,74858, HIP\,75132, HIP\,81284, and
HIP\,90864) have been analysed with MyGIsFOS. The other two stars
(HIP\,80124 and $\eta$\,UMi) have a high rotational velocity (\vsini )
and MyGIsFOS is not optimised for the analysis of these type of
objects.  For these two stars, we selected the features to take into
consideration in the analysis upon visual inspection.

MyGIsFOS does not derive \vsini, this is an input parameters.
We broaden the synthetic profiles taking into account the instrumental
broadening and the stellar rotation.
This approach is valid only with slow rotating stars, from our
experience below about 10\,\kms.
We optimised the analysis of MyGIsFOS by changing the total broadening for each star, and,
known the instrumental broadening, the \vsini\ we derive is in agreement,
within about one \kms\ with the values presented in Table\,\ref{obslog}.
For the two stars we did not analysed with MyGIsFOS (HIP\,80124 and $\eta$\,UMi),
by line profile fitting and assuming a resolving power of 40\,000, we derived
a \vsini\ of 26\,\kms\ and 85\,\kms, which are not in agreement with the
values provided in Table\,\ref{obslog}.

All the analysis was done with spectral synthesis computed with SYNTHE
\citep{synte93,kurucz05} in its Linux version \citep{ls04,ls05} based
on ATLAS\,9 and ATLAS\,12 \citep{kurucz93,kurucz05} models.

The stellar parameters derived are listed in Table\,\ref{param} whereas the
detailed chemical abundances are presented in tables on a star-by-star
basis and are discussed in their own sections.
The upper limits in the Li abundance evidence a non-detection of lithium.
It has been derived by applying the Cayrel's formula \citep{cayrel88} with
the measured  S/N
and assuming a 3\,$\sigma$ detection limit.

For testing purposes,
we analysed the spectrum of Ceres observed with SOPHIE during the same
observing run and the stellar parameters we derived are shown in
Table\,\ref{param}.  The reference solar abundances we adopted from
\citet{sunabbo} for Li, S, and Fe, and from \citet{lodders} for the
other elements.

\begin{table*}
\caption{Stellar parameters of the sample of stars; the first five stars were
analysed with the pipeline MyGIsFOS.}
\label{param}
\begin{tabular}{lrrrrrll}\hline
\hline
Star       &  \teff\  &   \logg\  & $\xi_{\rm turb}$  &  [Fe/H] & [$\alpha$/Fe] & \ion{Ca}{ii}\,-H,K &  $A$(Li)\\
\hline
 Ceres     &   5700   &  4.36 &  1.19 &--0.08 &  0.02  &              & \\
HIP\,67344 (HD\,120205)  &   5310   &  4.55 &  1.22 &  0.04 &  0.05  &     emission & $<0.54$\\ 
HIP\,74858 (HD\,136137)  &   5201   &  3.15 &  1.31 &  0.00 &  0.05  &     emission & $<0.45$\\ 
HIP\,75132 (HD\,136655)  &   5161   &  4.58 &  0.94 &  0.17  & 0.07   &    emission & $<0.49$ \\
HIP\,81284 (HD\,150202)  &   5198   &  3.26 &  1.14 &  0.04 &  0.09  &     emission & $1.37\pm 0.14$ \\
HIP\,90864 (HD\,171067)  &   5496   &  4.22 &  0.97 &--0.16 &  0.10  &              & $<0.74$\\ 
HIP\,80124 (HD\,147443)  &   6260   &  4.20 &  1.5  & +0.06 &  0.11  &     emission & $2.92\pm 0.09$\\
HIP\,79822 ($\eta$\,UMi) &   6946   &  4.05 & 2.0   &--0.10 & -0.04  &              & $<1.85$ \\ 
\hline
\end{tabular}
\end{table*}

\subsection{HIP\,67344 (HD\,120205)} 

\begin{figure*}
\includegraphics*[width=160mm]{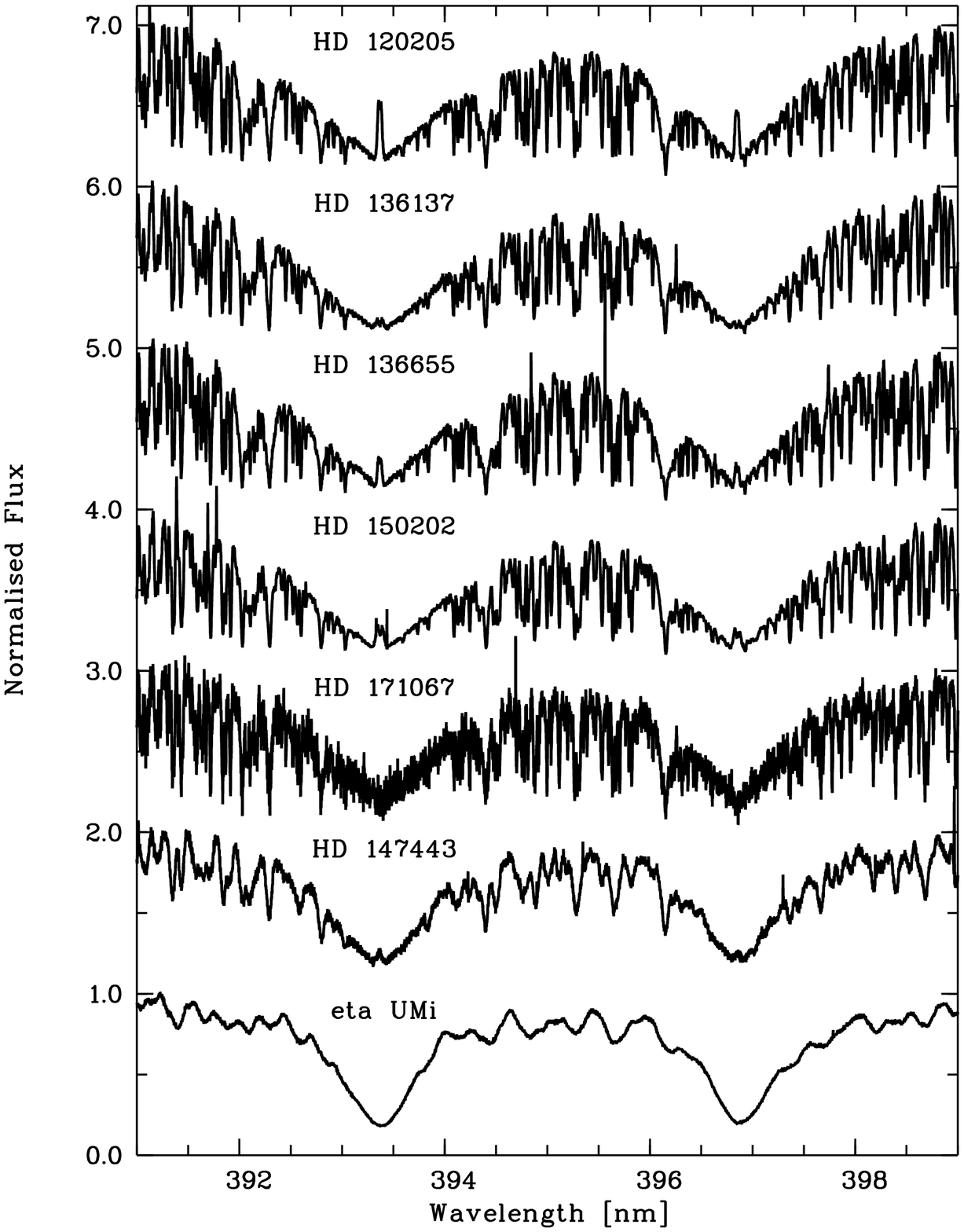}
\caption{The observed, normalised spectra in the region of the
  \ion{Ca}{ii}-H \& -K lines. For display reasons the spectra are
  vertically shifted.}
\label{cak}
\end{figure*}

This young, single G type star was analysed using MyGIsFOS. From
the SOPHIE spectrum, the following stellar parameters were determined:
(\teff/\logg/[Fe/H]) 5310\,K/4.55/+0.04 and a micro-turbulence
parameter of $\xi_{\rm turb}$=1.22\,\kms. The abundances are reported in
Table\,\ref{hd120205}. The star appears to be a dwarf star with
solar-metallicity.  The \ion{Ca}{ii}-H and -K lines show clear core
emission (see Fig.\,\ref{cak}).

The stellar parameters for this star available in the literature are
in good agreement with our results.  \citet{casagrande11}, who derive
\teff\ =5286\,K, give the most recent revision of the parameters
present in the Geneva-Copenhagen Catalogue.  The main improvement in
this edition is the use of the IR flux method to derive \teff\ and
this corresponds also to a new metallicity scale.  The effective
temperature from \citet{masana06} is $5317\pm 43$\,K.  From
\citet{mortier13} we have \teff\ =5303\,K, [Fe/H]=0.08, with a stellar
mass of $0.87\,{\rm M}_\odot$, and no detected planet.  From
\citet{strassmeier12} the stellar parameters are 5260\,K/4.36/+0.03, with
a $A$(Li)=1.13.  We could not derive the Li abundance from the SOPHIE
spectrum, because no Li line is detectable. On the other hand,
slightly blue-shifted in wavelength with respect to the Li doublet
position, we detect a line (at 670.7473\,nm) that is compatible with a
\ion{Sm}{ii} line, that would provide $\log{gf}\epsilon _{\rm Sm}=1.24$.

\begin{table*}
\caption{Chemical abundances of HIP\,67344 (HD\,120205) for \teff\ of
  5310\,K.}
\label{hd120205}
\begin{tabular}{lrrrrrrr}\hline
\hline
Element &   N$^a$  & $A_\odot$ &$A$(X)& [X/H] & $\sigma$ & [X/Fe] & $\sigma$ \\
\hline
\ion{Na}{i}  &   2 &  6.30 &  6.41 &  0.11 & 0.01   &  0.07 & 0.1080\\
\ion{Al}{i}  &   3 &  6.47 &  6.55 &  0.08 & 0.05   &  0.04 & 0.1203\\
\ion{Si}{i}  &   6 &  7.52 &  7.62 &  0.10 & 0.04   &  0.06 & 0.1155\\
\ion{Si}{ii} &   1 &  7.52 &  7.53 &  0.01 &        &--0.03 &       \\
\ion{S}{i}   &   1 &  7.16 &  7.23 &  0.07 &        &  0.03 &       \\
\ion{Ca}{i}  &   2 &  6.33 &  6.43 &  0.10 & 0.05   &  0.05 & 0.1174\\
\ion{Sc}{ii} &   6 &  3.10 &  3.21 &  0.11 & 0.13   &  0.06 & 0.1769\\
\ion{Ti}{i}  &  22 &  4.90 &  4.98 &  0.08 & 0.07   &  0.04 & 0.1261\\
\ion{Ti}{ii} &   4 &  4.90 &  5.04 &  0.14 & 0.09   &  0.10 & 0.1523\\
\ion{V}{i}   &  10 &  4.00 &  4.07 &  0.07 & 0.11   &  0.02 & 0.1561\\
\ion{Cr}{i}  &   6 &  5.64 &  5.68 &  0.04 & 0.06   &  0.00 & 0.1218\\
\ion{Mn}{i}  &   3 &  5.37 &  5.43 &  0.06 & 0.24   &  0.01 & 0.2605\\
\ion{Fe}{i}  &  72 &  7.52 &  7.56 &  0.04 & 0.11   &  0.00 & 0.1522\\
\ion{Fe}{ii} &  14 &  7.52 &  7.56 &  0.04 & 0.12   &  0.00 & 0.1695\\
\ion{Co}{i}  &   5 &  4.92 &  4.98 &  0.06 & 0.03   &  0.01 & 0.1113\\
\ion{Ni}{i}  &  14 &  6.23 &  6.21 &--0.02 & 0.17   &--0.06 & 0.2045\\
\ion{Cu}{i}  &   1 &  4.21 &  4.24 &  0.03 &        &--0.01 &       \\
\ion{Y}{ii}  &   3 &  2.21 &  2.22 &  0.01 & 0.21   &--0.03 & 0.2384\\
\hline
\multicolumn{8}{l}{$^a$ N: number of lines analysed of each element.}
\end{tabular}
\end{table*}

\subsection{HIP\,74858 (HD\,136137)}

We could not find any chemical analysis for this young evolved star in
the literature.  The \ion{Ca}{ii}-H and -K lines show a core emission
(see Fig.\,\ref{cak}).  According to our analysis based on
MyGIsFOS, the star is a sub-giant with solar metallicity. We derived
the stellar parameters 5201\,K/3.15/0.0. The chemical composition we
derived is reported in Table\,\ref{hd136137}.

\begin{table*}
\caption{Chemical abundances of HIP\,74858 (HD\,136137) for \teff\ of
  5201\,K.}
\label{hd136137}
\begin{tabular}{lrrrrrrr}\hline
\hline
Element &   N    & $A_\odot$ & $A$(X) & [X/H] & $\sigma$ & [X/Fe] & $\sigma$ \\
\hline
\ion{Na}{i}  &  2 &  6.30 &  6.49 &  0.19 & 0.01   &  0.19 & 0.1023  \\
\ion{Al}{i}  &  3 &  6.47 &  6.50 &  0.03 & 0.02   &  0.03 & 0.1039  \\
\ion{Si}{i}  &  7 &  7.52 &  7.53 &  0.01 & 0.04   &  0.02 & 0.1103  \\
\ion{Si}{ii} &  2 &  7.52 &  7.67 &  0.15 & 0.11   &  0.15 & 0.1646  \\
\ion{S}{i}   &  1 &  7.16 &  7.05 &--0.11 &        &--0.11 &         \\
\ion{Ca}{i}  &  3 &  6.33 &  6.37 &  0.04 & 0.03   &  0.05 & 0.1054  \\
\ion{Sc}{ii} &  4 &  3.10 &  3.26 &  0.16 & 0.11   &  0.16 & 0.1661  \\
\ion{Ti}{i}  & 22 &  4.90 &  4.96 &  0.06 & 0.10   &  0.06 & 0.1434  \\
\ion{Ti}{ii} &  5 &  4.90 &  4.90 &  0.00 & 0.20   &  0.00 & 0.2372  \\
\ion{V}{i}   &  9 &  4.00 &  4.01 &  0.01 & 0.12   &  0.02 & 0.1572  \\
\ion{Cr}{i}  &  6 &  5.64 &  5.62 &--0.02 & 0.04   &--0.02 & 0.1091  \\
\ion{Mn}{i}  &  3 &  5.37 &  5.34 &--0.03 & 0.27   &--0.03 & 0.2871  \\
\ion{Fe}{i}  & 77 &  7.52 &  7.52 &  0.00 & 0.10   &  0.00 & 0.1445  \\
\ion{Fe}{ii} & 16 &  7.52 &  7.52 &  0.00 & 0.12   &  0.00 & 0.1729  \\
\ion{Co}{i}  &  4 &  4.92 &  4.91 &--0.01 & 0.08   &--0.01 & 0.1276  \\
\ion{Ni}{i}  & 14 &  6.23 &  6.19 &--0.04 & 0.13   &--0.04 & 0.1691  \\
\ion{Cu}{i}  &  1 &  4.21 &  4.13 &--0.08 &        &--0.08 &         \\
\ion{Y}{ii}  &  3 &  2.21 &  2.35 &  0.14 & 0.23   &  0.14 & 0.2633  \\
\hline
\end{tabular}
\end{table*}

\subsection{HIP\,75132 (HD\,136655)}

This star is in a double system (separation 69\,arcsec) with the
companion (K\,2) almost two magnitudes fainter (V=\,9.02 and 10.99,
respectively), with a high probability to be a physical double system
\citep{shaya11}.

The stellar parameters we derive for this star (5161\,K/4.58/+0.17) are
slightly different with respect to those in \citet{strassmeier12}
(5010\,K/4.27/+0.09).  In particular we derived an effective
temperature larger by 151\,K, a gravity 0.31\,dex higher, and a metallicity
0.08\,dex higher. If we fix the temperature
at 5010\,K from \citet{strassmeier12}, we obtain a perfect agreement
with \citet{strassmeier12}.  Our chemical analysis, reported in
Table\,\ref{hd136655}, shows that this star has an over-solar
metallicity, with a low Mn abundance, that can probably
be explained with effect due to deviation from local thermodynamical
equilibrium (LTE).

The radial velocity of this star is considered to be constant by
\citet{Griffin2010} $-31.03\pm0.06$\,\kms\ from 9 CORAVEL measurements
between March and September 2009. The constancy of the radial velocity
is confirmed by \citet{strassmeier12} who find $-31.921\pm 0.035$\,\kms\
from 112 STELLA measurements that span 842 days.  Our SOPHIE radial
velocity is slightly higher than both these series of measurements,
(0.4\,\kms\ larger than the STELLA measurement and 1.3\,\kms\ than the
CORAVEL measurement).  STELLA radial velocities are offset to the
CORAVEL RVs by $+0.503$\,\kms\ \citep{strassmeier12}.  We do not
believe this points towards any radial velocity variability of this
star, but rather to zero point offsets between the three
instruments. See for example \citet{Pasquini} for combining SOPHIE
radial velocities with those of other spectrographs.

\begin{table*}
\caption{Chemical abundances of HIP\,75132 (HD\,136655) for \teff\ of
  5161\,K.}
\label{hd136655}
\begin{tabular}{lrrrrrrr}\hline
\hline
Element &   N    & $A_\odot$ & $A$(X) & [X/H] & $\sigma$ & [X/Fe] & $\sigma$ \\
\hline
\ion{Al}{i}  &   2 &  6.47 &  6.71 &  0.24 & 0.12   &  0.07 & 0.1701 \\
\ion{Si}{i}  &   6 &  7.52 &  7.86 &  0.34 & 0.15   &  0.17 & 0.1880 \\
\ion{Ca}{i}  &   2 &  6.33 &  6.57 &  0.24 & 0.02   &  0.07 & 0.1187 \\
\ion{Sc}{ii} &   7 &  3.10 &  3.43 &  0.33 & 0.15   &  0.15 & 0.2274 \\
\ion{Ti}{i}  &  18 &  4.90 &  5.16 &  0.26 & 0.08   &  0.09 & 0.1438 \\
\ion{Ti}{ii} &   4 &  4.90 &  5.17 &  0.27 & 0.17   &  0.09 & 0.2357 \\
\ion{V}{i}   &   6 &  4.00 &  4.28 &  0.28 & 0.22   &  0.11 & 0.2483 \\
\ion{Cr}{i}  &   5 &  5.64 &  5.87 &  0.23 & 0.06   &  0.06 & 0.1328 \\
\ion{Mn}{i}  &   1 &  5.37 &  5.39 &  0.02 &        &--0.15 &        \\
\ion{Fe}{i}  &  65 &  7.52 &  7.69 &  0.17 & 0.12   &  0.00 & 0.1663 \\
\ion{Fe}{ii} &  15 &  7.52 &  7.70 &  0.18 & 0.17   &  0.00 & 0.2370 \\
\ion{Co}{i}  &   3 &  4.92 &  5.28 &  0.36 & 0.11   &  0.19 & 0.1579 \\
\ion{Ni}{i}  &  13 &  6.23 &  6.49 &  0.26 & 0.18   &  0.09 & 0.2173 \\
\ion{Cu}{i}  &   1 &  4.21 &  4.57 &  0.36 &        &  0.19 &        \\
\ion{Y}{ii}  &   3 &  2.21 &  2.25 &  0.04 & 0.21   &--0.13 & 0.2687 \\
\ion{Ba}{ii} &   1 &  2.17 &  2.30 &  0.13 &        &--0.05 &        \\
\hline
\end{tabular}
\end{table*}

\subsection{HIP\,81284 (HD\,150202)}\label{sec_hd150202}

This star is known in the literature to be an SB1 system
\citep{strassmeier00} and as such it was studied by \citet{griffin09}.
The late spectral type K0 quoted by \citet{griffin09} is taken from
the Chromospheric Active Binaries catalogue by \citet{eker08} who gave
a \teff\ of 4950\,K and a \vsini\ of 8.1\,\kms, values taken from
\citet{strassmeier00}. However, it is not mentioned that these authors
measured two spectra from which they derived two different values of
\vsini, 5.6 and 8.1\,\kms.

From our SOPHIE spectrum we derive an intermediate value \vsini\,$\sim 7$\,\kms.
The spectral type according to \citet{strassmeier12}
should in fact be G8III.  This star is a photometric variable, with a
light curve in Hipparcos magnitude from ${\rm Hp}_{\rm max}=8.09$ to
${\rm Hp}_{\rm min}=8.16$ giving 8.099 and 8.140 V magnitude,
respectively.  The photometric variability, discovered by Hipparcos,
satisfies the criteria of GCVS (General Catalogue of Variable Stars,
\citealt{GCVS08,GCVS10}) and as such this star has the designation of
GI Dra \citep{kazarovets99}.  The assigned variability type is
``SRD:'' i.e. uncertain Semi-Regular variable of type D.  Semi-regular
variable are giants and super-giants stars of spectral type F, G, or
K, sometimes with emission lines in their spectra. The amplitudes of
light variation are in the range between 0.1 and 4\,mag, and the range
of periods is between 29 and 1100\,days (SX Her, SV UMa).

\citet{strassmeier12} show this star to be active, and we confirm it by
seeing a clear emission in the core of the \ion{Ca}{ii}-H and -K lines
(see Fig.\,\ref{cak}).  Our radial velocity of about
$-27$\,\kms\ puts the star close to the minimum of the
\vrad\ variation which ranges from $-30$\,\kms\ to
$+17$\,\kms\ \citep{griffin09}.

A determination of effective temperature, gravity and metallicity for
this star, based on selected spectral orders in the range 549-623\,nm,
has previously been done by \citet{strassmeier12} who determined a
\teff\ of 5010\,K, \logg\ =3.06, and $[{\rm Fe/H}]=-0.12$.
We derive a \teff\ of 5198\,K, 188\,K higher than \citet{strassmeier12},
\logg\ = 3.26, and $[{\rm Fe/H}]=0.04$ (Table\,\ref{param}).
\citet{mcdonald12} used the spectral energy distribution from 420\,nm
to 25000\,nm to derive \teff, luminosity, and IR excess of a large
sample of Hipparcos stars. They obtained a \teff\ of 5100\,K,
$L=2.72 {\rm L_\odot}$ and an average IR excess of 1.330 with a peak
at 12\,nm for HIP\,81284.

By fixing the temperature at 5010\,K, as in \citet{strassmeier12}, we
have a reasonable agreement of the stellar parameters
from our analysis, 5010\,K/2.72/0.0, versus 5010\,K/3.06/--0.12 from
\citep{strassmeier12}.

Lithium can be detected from the 670.8\,nm resonance line and we performed
a full 3D NLTE spectral analysis of the lithium doublet with the
\nlte+\linfor\ codes, based on 3D hydro-dynamical stellar atmospheres
taken from the CIFIST 3D model atmosphere grid \citep{ludwig09}. These
models are the outcome of realistic numerical simulations performed
with the \COBOLD\ code \citep{freytag12}.  From our high-resolution
SOPHIE spectrum (R$\sim$40\,000, SNR$\sim$80) it is clear that the
Li doublet at 670.8\,nm is blended with lines of other elements.
For this reason the spectral synthesis was carried out by using the
list of blending lines taken from \citet{melendez12} in the range
670.70\,nm and 670.86\,nm, as presented in Table\,\ref{Melendez}.
To compute the lithium line profile to an adequate accuracy, we considered
its full isotopic and hyper-fine structure computed by
Kurucz\footnote{http://kurucz.harvard.edu/atoms/0300/lidlines.dat},
represented by twelve components, six of which belong to $^{6}$Li
and six to $^{7}$Li, as shown in Table \ref{Lithium_HFS}.

\begin{table}
\caption{List of atomic and molecular data of \citet{melendez12} for
the blending lines in the region around 670.8\,nm.}
\label{Melendez}
\begin{tabular}{cccc}\hline
\hline
$\lambda$ [nm]& Element/Molecule & $\chi$ [eV] & $\log{gf}$\\
\hline
670.7000 & \ion{Si}{i}     & 5.954 & -2.560 \\
670.7172 & \ion{Fe}{i}     & 5.538 & -2.810 \\
670.7205 & \ion{CN}{}      & 1.970 & -1.222 \\
670.7272 & \ion{CN}{}      & 2.177 & -1.416 \\
670.7282 & \ion{CN}{}      & 2.055 & -1.349 \\
670.7300 & \ion{C$_{2}$}{} & 0.933 & -1.717 \\
670.7371 & \ion{CN}{} 	   & 3.050 & -0.522 \\
670.7433 & \ion{Fe}{i}     & 4.608 & -2.250 \\
670.7460 & \ion{CN}{}      & 0.788 & -3.094 \\
670.7461 & \ion{CN}{}      & 0.542 & -3.730 \\
670.7470 & \ion{CN}{}      & 1.880 & -1.581 \\
670.7473 & \ion{Sm}{ii}    & 0.933 & -1.910 \\
670.7548 & \ion{CN}{}      & 0.946 & -1.588 \\
670.7595 & \ion{CN}{}      & 1.890 & -1.451 \\
670.7596 & \ion{Cr}{i}     & 4.208 & -2.667 \\
670.7645 & \ion{CN}{}      & 0.946 & -3.330 \\
670.7660 & \ion{C$_{2}$}{} & 0.926 & -1.743 \\
670.7809 & \ion{CN}{}      & 1.221 & -1.935 \\
670.7848 & \ion{CN}{}      & 3.600 & -2.417 \\
670.7899 & \ion{CN}{}      & 3.360 & -3.110 \\
670.7930 & \ion{CN}{}      & 1.980 & -1.651 \\
670.7970 & \ion{C$_{2}$}{} & 0.920 & -1.771 \\
670.7980 & \ion{CN}{}      & 2.372 & -3.527 \\
670.8023 & \ion{Si}{i}     & 6.000 & -2.800 \\
670.8026 & \ion{CN}{}      & 1.980 & -2.031 \\
670.8094 & \ion{V}{i}      & 1.218 & -2.922 \\
670.8099 & \ion{Ce}{ii}    & 0.701 & -2.120 \\
670.8147 & \ion{CN}{}      & 1.870 & -1.884 \\
670.8282 & \ion{Fe}{i}     & 4.988 & -2.700 \\
670.8315 & \ion{CN}{}      & 2.640 & -1.719 \\
670.8347 & \ion{Fe}{i}     & 5.486 & -2.580 \\
670.8370 & \ion{CN}{}      & 2.640 & -2.540 \\
670.8420 & \ion{CN}{}      & 0.768 & -3.358 \\
670.8534 & \ion{Fe}{i}     & 5.558 & -2.936 \\
670.8541 & \ion{CN}{}      & 2.500 & -1.876 \\
670.8577 & \ion{Fe}{i}     & 5.446 & -2.684 \\
\hline
\end{tabular}
\end{table}

\begin{table}
\caption{Hyper-fine structure of the lithium resonance line.}
\label{Lithium_HFS}
\begin{tabular}{cccc}\hline
\hline
$\lambda$ (\AA) & Li isotope & $\chi$ [eV] & $\log{gf}$\\
\hline
670.7756 & $^{7}$Li & 0.000 & -0.428 \\
670.7768 & $^{7}$Li & 0.000 & -0.206 \\
670.7907 & $^{7}$Li & 0.000 & -0.808 \\
670.7908 & $^{7}$Li & 0.000 & -1.507 \\
670.7919 & $^{7}$Li & 0.000 & -0.808 \\
670.7920 & $^{7}$Li & 0.000 & -0.808 \\
670.7920 & $^{6}$Li & 0.000 & -0.479 \\
670.7923 & $^{6}$Li & 0.000 & -0.178 \\
670.8069 & $^{6}$Li & 0.000 & -0.831 \\
670.8070 & $^{6}$Li & 0.000 & -1.734 \\
670.8074 & $^{6}$Li & 0.000 & -0.734 \\
670.8075 & $^{6}$Li & 0.000 & -0.831 \\
\hline
\end{tabular}
\end{table}

The lithium abundance was derived through fitting a grid of synthetic
spectra to the SOPHIE data. Three parameters were varied independently
to find the best fit ($\chi^2$ minimisation): the total Li abundance
$A({\rm Li})$, which controls the line strength, a global wavelength shift
$\Delta v$, and a residual line broadening described by a Gaussian
kernel with FWHM $V_{\rm BR}$ (in velocity space). During the fitting
procedure we kept the continuum fixed, and since the lithium isotopic
ratio $^6{\rm Li}/^7{\rm Li}$ was not the goal of our analysis for this
star, we constrained its value to the solar value of
$\sim 8$\% \citep{Lemoine1995}, which is well matched by the meteoritic
value of 8.33\% \citep{anders1989}.

\begin{figure}
\begin{minipage}[c][10.8cm][t]{\columnwidth}
  \centering
  \includegraphics*[width=80mm]{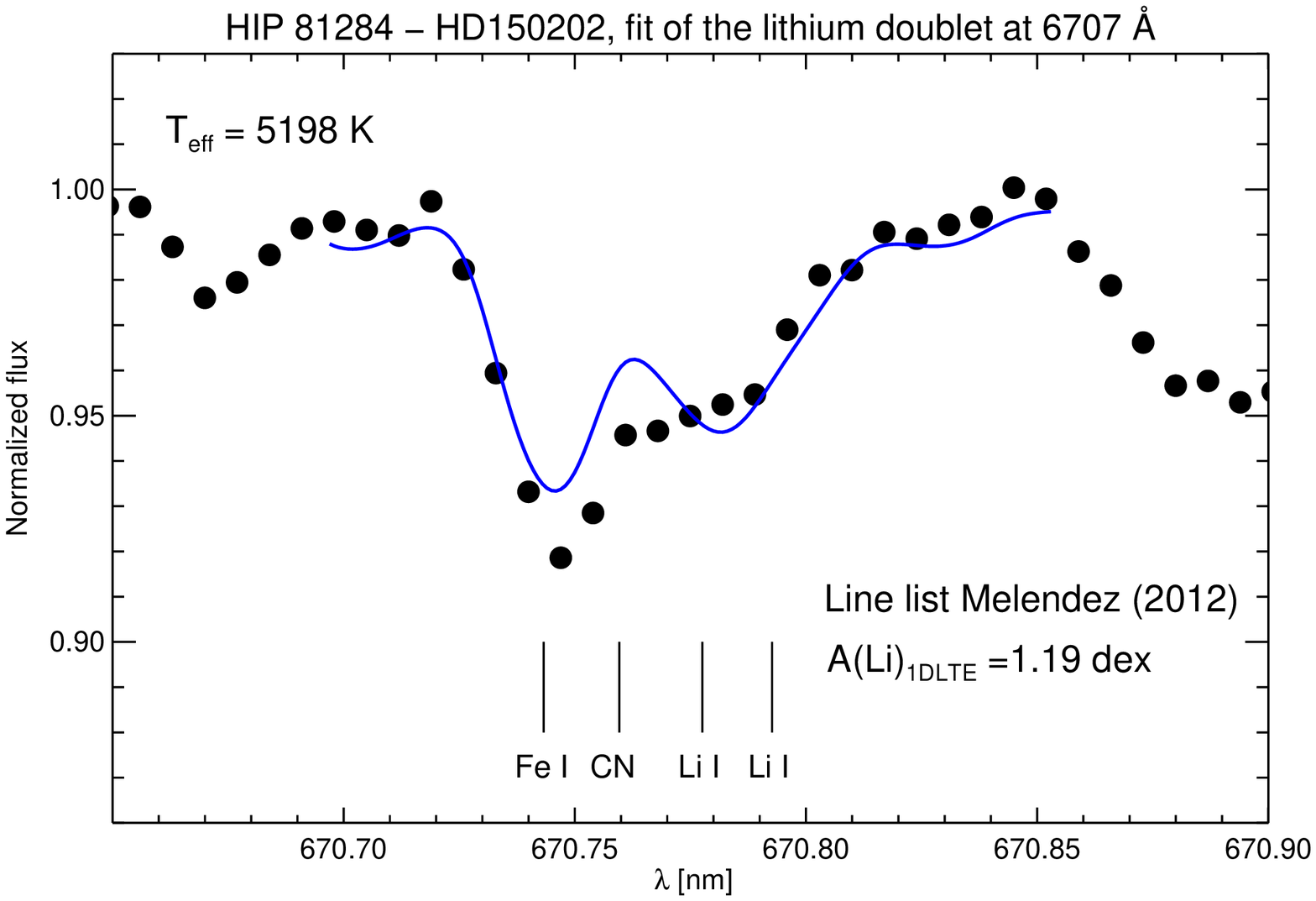}
  \includegraphics*[width=80mm]{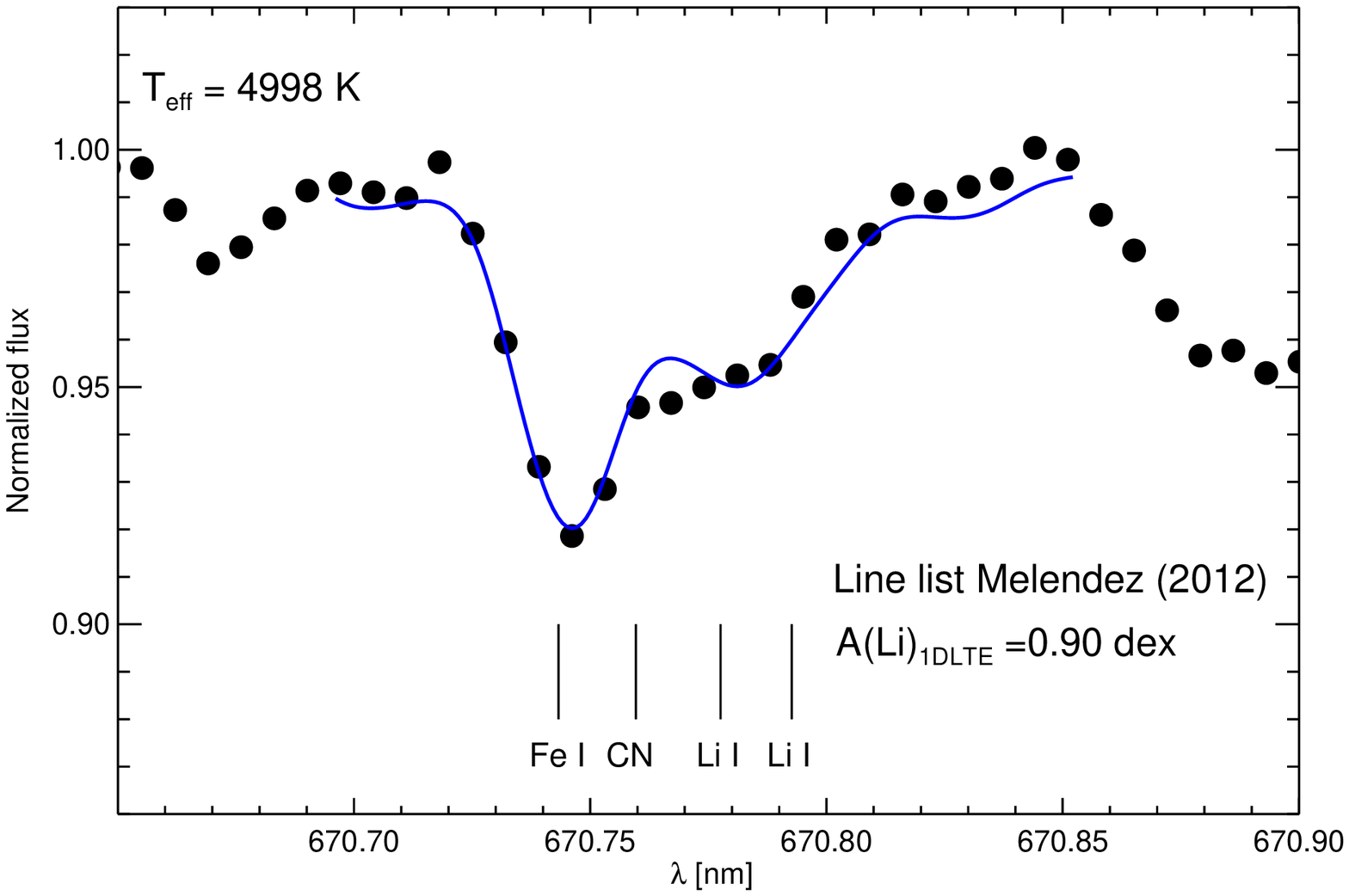}
\end{minipage}
\caption{The best-fit synthetic data (solid lines) to the \ion{Li}{i} 670.8\,nm
region from the observed SOPHIE spectrum of HIP\,81284 (HD\,150202)
(black dots). Computations were done in LTE with two different 1D ATLAS9 models
assuming \teff=5198\,K (upper panel) and \teff=4998\,K (lower panel).
The abundances resulting from the fits are $A({\rm Li})_{\rm  1DLTE}=1.19$
for \teff=5198\,K, and $A({\rm Li})_{\rm 1DLTE}=0.90$ for \teff=4998\,K.
A cooler temperature seems more suitable for achieving a good fit
of the blending lines near the lithium doublet (mostly \ion{Fe}{i} and
\ion{CN}{}).}
\label{fitsHD150202}
\end{figure}

We started by using synthetic spectra computed in LTE from a standard 1D
ATLAS9 model atmosphere \citep{kurucz93,kurucz05} with parameters
\teff =5198\,K, \logg\,=3.26, and [Fe/H]=0.0, and we find
$A({\rm Li})_{\rm 1DLTE-T5198 K}=1.19$. Subsequently, the abundance
resulting from the fit (shown in Fig.\,\ref{fitsHD150202}, upper panel) is
corrected for 3D hydro-dynamical+NLTE effects.

Following the procedure outlined in Appendix \ref{sec:A}, we derive a
3D+NLTE correction of $\Delta_{\rm 3DNLTE} = +0.18$\,dex at \teff =5198\,K,
resulting in $A({\rm Li})_{\rm 3DNLTE-T5198 K} = 1.37\pm0.14\,{\rm dex}$. This result is
in very good agreement with $A({\rm Li})=1.32\pm0.16\,{\rm dex}$\,dex from
\citet{strassmeier12} -- a value obtained from equivalent-width
measurements and NLTE corrected using the computations of
\citet{pavlenko96}, even though these authors adopted a lower effective
temperature (see below). The error associated with $A({\rm Li})$ was calculated by
examining both systematic and statistical uncertainties.  The
systematic error arises from the uncertainty of the stellar parameters,
and it was tested by means of ATLAS9 models under the assumption of LTE.
Since the lithium line is mainly sensitive to the effective temperature
of the star, the error due to changes in gravity and metallicity are not
considered. We employed 1D ATLAS9 model
atmospheres with \logg\,=3.26, [Fe/H]=0.0, and $T_{\rm eff}=5198\pm 100$\,K.
A change in temperature of $\pm 100$\,K is
translated into a change in $A({\rm Li})$ of $_{-0.14}^{+0.13}$\,dex.  The
statistical error comprises the global uncertainty resulting from the
fit between observed data and our synthetic spectra. The final error
on $A({\rm Li})$ of $0.14$\,dex is the sum in quadrature of the two
contributions, where the change in $A({\rm Li})$ due to changes in
\teff\ is clearly predominant.

We also performed a 3D+NLTE spectrum analysis considering
the lower \teff$\sim$5010\,K of \citet{strassmeier12}, following the
same procedure.
We find $A({\rm Li})_{\rm 1DLTE-T4998 K}$\,=\,$0.90$\,dex which was
corrected for the 3D hydro-dynamical effects to get the final value of
$A({\rm Li})_{\rm 3DNLTE-T4998 K}=1.10\pm0.15$\,dex,  almost a factor
two lower than for \teff$\,=\,$5198\,K.
Assuming a \teff\ closer to the one from \citet{strassmeier12}
(Fig.\,\ref{fitsHD150202}, lower panel) allows to achieve a better fit of
the observed \ion{Fe}{i} and \ion{CN}{} lines that mostly contribute to the
blending of the lithium resonance doublet. The higher effective temperature
\teff=5198\,K derived from our high-resolution SOPHIE spectrum seems too
high to fit properly the main features responsible for the blending
(Fig.\,\ref{fitsHD150202}, upper panel).

The results of the analysis of the other elements present in the
spectrum are shown in Table\,\ref{hd150202}.

\begin{table*}
\caption{Chemical abundances of HIP\,81284 (HD\,150202) for \teff\ of 5198\,K.}
\label{hd150202}
\begin{tabular}{lrrrrrrr}\hline
\hline
Element &   N & $A_\odot$ & $A$(X) & [X/H] & $\sigma$ & [X/Fe] & $\sigma$ \\
\hline
\ion{Li}{i}  &  2 &  1.03 &  1.37 &       & 	   &       &        \\
\ion{Na}{i}  &  2 &  6.30 &  6.52 &  0.22 & 0.02   &  0.18 & 0.1065 \\
\ion{Al}{i}  &  3 &  6.47 &  6.54 &  0.07 & 0.02   &  0.03 & 0.1078 \\
\ion{Si}{i}  &  7 &  7.52 &  7.56 &  0.04 & 0.07   &--0.01 & 0.1237 \\
\ion{Si}{ii} &  1 &  7.52 &  7.71 &  0.19 &        &  0.14 &        \\
\ion{S}{i}   &  1 &  7.16 &  7.08 &--0.08 &        &--0.12 &        \\
\ion{Ca}{i}  &  3 &  6.33 &  6.46 &  0.13 & 0.02   &  0.09 & 0.1071 \\
\ion{Sc}{ii} &  5 &  3.10 &  3.26 &  0.16 & 0.10   &  0.11 & 0.1701 \\
\ion{Ti}{i}  & 21 &  4.90 &  5.05 &  0.15 & 0.12   &  0.11 & 0.1588 \\
\ion{Ti}{ii} &  6 &  4.90 &  4.92 &  0.02 & 0.23   &--0.03 & 0.2678 \\
\ion{V}{i}   &  8 &  4.00 &  4.11 &  0.11 & 0.13   &  0.07 & 0.1654 \\
\ion{Cr}{i}  &  6 &  5.64 &  5.72 &  0.08 & 0.07   &  0.04 & 0.1238 \\
\ion{Fe}{i}  & 78 &  7.52 &  7.56 &  0.04 & 0.11   &  0.00 & 0.1488 \\
\ion{Fe}{ii} & 16 &  7.52 &  7.57 &  0.05 & 0.14   &  0.00 & 0.1940 \\
\ion{Co}{i}  &  3 &  4.92 &  4.96 &  0.04 & 0.12   &--0.01 & 0.1577 \\
\ion{Ni}{i}  & 14 &  6.23 &  6.26 &  0.03 & 0.14   &--0.01 & 0.1714 \\
\ion{Cu}{i}  &  1 &  4.21 &  4.26 &  0.05 &        &  0.00 &        \\
\ion{Y}{ii}  &  4 &  2.21 &  2.30 &  0.09 & 0.19   &  0.04 & 0.2367 \\
\hline
\end{tabular}
\end{table*}

\subsection{HIP\,90864 (HD\,171067)}

This star has already been analysed by several authors.  Seven
\teff\ values are reported in the Pastel catalogue \citep{soubiran10}
and they are in very good agreement ($5625\pm 55$) among themselves
and in even better agreement ($5646\pm 25$) if we remove the coolest
value of 5500\,K from \citet{pompeia11}.  This latter value is in good
agreement with the \teff\ we derive (5496\,K) and the value of 5520\,K
from \citet{strassmeier12}.  Our relatively low [Fe/H]=--0.16, is
lower than the three values in the Pastel catalogue ($-0.03\pm 0.01$)
but again in agreement with [Fe/H]=--0.15 from \citet{strassmeier12}.
Our chemical analysis is presented in Table\,\ref{hd171067}.
\citet{valenti05} analysed a Keck spectrum and derived the stellar
parameters (\teff\ =5643\,K, \logg\ =4.50, [M/H]=--0.03,
\vsini\ =1.2\,\kms) and the abundances of $[{\rm Na/H}]=-0.06$, $[{\rm
    Si/H}]=-0.04$, $[{\rm Ti/H}]=-0.05$, $[{\rm Fe/H}]=-0.03$, and
$[{\rm Ni/H}]=-0.06$.  Our values are lower, but this is not
surprising since out effective temperature is about 150\,K lower than
their adopted value.

\begin{table*}
\caption{Chemical abundances of HIP\,90864 (HD\,171067) for \teff\ of 5496\,K..}
\label{hd171067}
\begin{tabular}{lrrrrrrr}\hline
\hline
Element &   N    & $A_\odot$ & $A$(X) & [X/H] & $\sigma$ & [X/Fe] & $\sigma$ \\
\hline
\ion{Na}{i}  &    2 &  6.30 &  6.14 &--0.16 & 0.02   &  0.01 & 0.1091 \\
\ion{Al}{i}  &    3 &  6.47 &  6.29 &--0.18 & 0.06   &--0.01 & 0.1243 \\
\ion{Si}{i}  &    4 &  7.52 &  7.41 &--0.11 & 0.08   &  0.06 & 0.1323 \\
\ion{Si}{ii} &    2 &  7.52 &  7.61 &  0.09 & 0.18   &  0.25 & 0.2125 \\
\ion{S}{i}   &    1 &  7.16 &  7.03 &--0.13 &        &  0.04 &        \\
\ion{Ca}{i}  &    6 &  6.33 &  6.26 &--0.07 & 0.06   &  0.10 & 0.1243 \\
\ion{Sc}{ii} &    6 &  3.10 &  2.97 &--0.13 & 0.10   &  0.03 & 0.1462 \\
\ion{Ti}{i}  &   25 &  4.90 &  4.70 &--0.20 & 0.05   &--0.03 & 0.1170 \\
\ion{Ti}{ii} &    7 &  4.90 &  4.71 &--0.19 & 0.23   &--0.03 & 0.2542 \\
\ion{V}{i}   &    8 &  4.00 &  3.68 &--0.32 & 0.09   &--0.16 & 0.1431 \\
\ion{Cr}{i}  &    5 &  5.64 &  5.45 &--0.19 & 0.03   &--0.02 & 0.1131 \\
\ion{Mn}{i}  &    8 &  5.37 &  5.27 &--0.10 & 0.24   &  0.07 & 0.2668 \\
\ion{Fe}{i}  &  107 &  7.52 &  7.36 &--0.16 & 0.11   &  0.00 & 0.1524 \\
\ion{Fe}{ii} &   14 &  7.52 &  7.36 &--0.16 & 0.11   &  0.00 & 0.1508 \\
\ion{Co}{i}  &    6 &  4.92 &  4.69 &--0.23 & 0.09   &--0.06 & 0.1391 \\
\ion{Ni}{i}  &   16 &  6.23 &  6.05 &--0.18 & 0.20   &--0.01 & 0.2254 \\
\ion{Cu}{i}  &    2 &  4.21 &  3.95 &--0.26 & 0.07   &--0.10 & 0.1310 \\
\ion{Y}{ii}  &    4 &  2.21 &  2.01 &--0.20 & 0.16   &--0.04 & 0.1894 \\
\ion{Ba}{ii} &    1 &  2.17 &  2.30 &  0.13 &        &  0.29 &        \\
\hline
\end{tabular}
\end{table*}

\subsection{HIP\,80124 (HD\,147443)} \label{sec_hip80124}

HIP\,80124 is one of the six potential solar siblings selected by
\citet{brown10}.  A solar sibling is a star that has the same age and
chemical composition of the Sun, because it formed at the same time
and from the same cluster.  It is not necessarily a solar analog. That
requires the star to have the same \teff, \logg, mass, and luminosity
as the Sun.

\citet{brown10} provides a radial velocity of $8.1\pm 0.2$\,\kms\ for
this star.  \citet{batista14} conducted an abundance analysis of a
large sample of solar siblings, but they excluded HIP\,80124 because its
\vsini\ was too high: roughly 26\,\kms\ as we derived from the SOPHIE
spectrum.  The radial velocity of the two spectra taken on 2nd and 3rd
March 2014 are substantially different; $-6.5426\pm 0.0077$\,\kms\ and
$+4.1704\pm0.0134$\,\kms, respectively.  The star is metal-rich and
shows signs of activity (see Fig.\,\ref{cak}).  We derived an effective
temperature of 6260\,K from the wings of H$\alpha$, which is in good
agreement with the effective temperature of 6194\,K from
\citet{casagrande11}.  We can associate an uncertainty of 100\,K to
\teff.  The logarithmic surface gravity of 4.2 ([cgs] units) was derived from the
Hipparcos parallax, assuming 1.2\,${\rm M}_\odot$.  An uncertainty of
0.2\,dex was used for \logg\ as a consequence of poor stellar mass and
parallax determinations.  \cite{casagrande11} report 4.21 in surface
gravity.  It was not possible to perform an automatic chemical
analysis with MyGIsFOS for HIP\,80124, as was done for the other
stars, because the high rotation of the star is responsible for the excessive blending of
spectral lines.  We selected reasonably unblended lines and measured the
equivalent widths, and for blended lines we performed a line profile
fitting to derive the abundances.  The results are presented in
Table\,\ref{abbohip80124}.

\begin{table}
\caption{Chemical abundances of HIP\,80124 (HD\,147443) for \teff\ of 6260\,K.}
\label{abbohip80124}
\begin{tabular}{lrrrrrr}\hline
\hline
Element &   N    & $A_\odot$ & $A$(X) & [X/H] & $\sigma$ & [X/Fe] \\
\hline
\ion{Li}{i}  &    1 &  1.03 &  2.92 &       & 0.10   &        \\
\ion{Na}{i}  &    1 &  6.30 &  6.45 & +0.15 & 0.10   & +0.09  \\
\ion{Si}{i}  &    4 &  7.52 &  7.52 & +0.00 & 0.15   &--0.06  \\
\ion{Si}{ii} &    2 &  7.52 &  7.62 & +0.10 & 0.04   & +0.04  \\
\ion{S}{i}   &    1 &  7.16 &  7.04 &--0.08 & 0.15   &--0.14  \\
\ion{Ca}{i}  &    4 &  6.33 &  6.50 & +0.17 & 0.18   & +0.11  \\
\ion{Ti}{i}  &    1 &  4.90 &  4.91 & +0.01 & 0.20   &--0.05  \\
\ion{Fe}{i}  &   44 &  7.52 &  7.58 & +0.06 & 0.10   &  0.00  \\
\ion{Fe}{ii} &    3 &  7.52 &  7.71 & +0.19 & 0.25   &  0.00  \\
\ion{Ni}{i}  &    5 &  6.23 &  6.32 & +0.09 & 0.06   & +0.03  \\
\ion{Ba}{ii} &    1 &  2.17 &  2.20 & +0.03 & 0.15   &--0.03  \\
\hline
\end{tabular}
\end{table}

A strong \ion{Li}{i} doublet at 670.8\,nm is visible in the observed
spectrum, with an equivalent width of 8.7\,pm (87\,m\AA).  As for HIP\,81284
(HD\,150202), we derived $A({\rm Li})$ by fitting the observed data with a
grid of synthetic spectra computed with a 1D ATLAS9 model atmosphere,
and subsequently considered 3D effects and departures from LTE conditions
through the application of ``3D+NLTE'' abundance corrections
(see Appendix \ref{sec:A} for the procedure).

\begin{figure}
\includegraphics*[width=80mm]{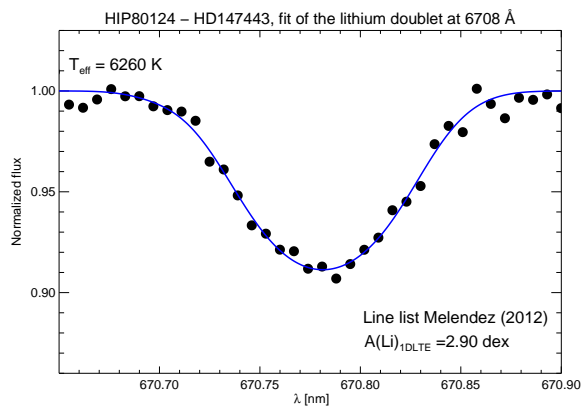}
\caption{The result of the best-fitting to the \ion{Li}{i} 670.8\,nm
 region (solid line) of the SOPHIE spectrum of HIP\,80124 (HD\,147443)
(black dots). We used a 1D ATLAS9 model in LTE with \teff=6260\,K
 as explained in Section \ref{sec_hip80124}. We find
$A({\rm Li})_{\rm 1DNLTE-T6260 K}=2.90$, which was corrected for
3D hydro-dynamical and non-LTE effects to get the final
value ${A}({\rm  Li})_{\rm 3DNLTE-T6260 K}=2.92\pm0.09$.}
\label{hip80124fit}
\end{figure}

We performed our spectral analysis including the list of atomic and
molecular data of \citet{melendez12} (see Table\,\ref{Melendez}) and
derive the final value of
${A}({\rm Li})_{\rm 3DNLTE-T6260 K}=2.92\pm0.09\,{\rm dex}$; the outcome of the
fitting procedure is shown in Fig.\,\ref{hip80124fit}.
We computed the uncertainties as done for HIP\,81284 (HD\,150202): a
change in \teff\ of $\pm 100$\,K resulted in a change in $A({\rm Li})$ by
$_{-0.08}^{+0.09}$\,dex.

Since a 3D model is available at the right stellar parameters for this
object, we can also fit the observed spectrum directly with the grid of
3D synthetic spectra. The result is basically identical to the above abundance,
${A}({\rm Li})_{\rm 3DNLTE-T6260 K}=2.89$, demonstrating the internal
consistency of our 3D correction procedure.

To understand if this star is a solar sibling, we compared its
chemical composition reported in Table \ref{abbohip80124} with
the reference solar abundances \citep{sunabbo,lodders}.
We did not find a good agreement.  We found that HIP\,80124 (HD\,147443)
has a higher Li abundance ($A({\rm Li})=2.92\pm 0.09$) compared to the
Sun ($A({\rm Li})_\odot$=1.03$\pm$0.03, \citealt{sunabbo}).  This by itself
is not surprising since the star is much warmer than the Sun, thus
convection should not be as effective in depleting lithium in its
atmosphere. However, if we consider the lithium abundances in the open
cluster M\,67 \citep{Pasquini08}, that has the same age and chemical
composition as the Sun, we see that stars of roughly the same
temperature as HIP\,80124 (HD\,147443) have $A({\rm Li})\approx 2.5$ (see
Figure 6 of \citealt{Pasquini08}). This suggests that this star is
younger than the Sun (and M\,67).
The logarithmic lithium abundance for 12 G-dwarf stars in the 800-Myr
Open Cluster IC4756 is 2.39$\pm$0.17 \citep{strassmeier15}.
This suggests that
HIP\,80124 (HD\,147443) may even be younger than that.
The two measured RVs suggest that the
star could be  a close binary, since we do not detect the spectrum
of a companion star, it is likely that it is much less luminous.
Further observations of this star are encouraged to study its
radial velocity variations.
On the assumption that M\,67 and
the Sun formed with the same lithium abundance, equal to the
meteoritic abundance ($A({\rm Li})=3.28$, \citealt{lodders}), the lower Li
abundance observed in M\,67 suggests that some phenomenon (for example
diffusion) has decreased the Li abundance even in the hottest stars.
Since this phenomenon should be time-dependent we expect that a
younger star, that started with the same Li abundance, should have a
higher Li abundance.
In addition a younger star would be born at a time when the
interstellar medium has been further enriched in Li, thus its initial
lithium abundance might have been even higher than the
meteoritic value. Both effects imply that a younger star should have a
higher lithium abundance.  The high lithium abundance for such a
rapidly rotating star is not uncommon among active stars.  For an
inclination angle between 30$^\circ$ and 90$^\circ$ the
\vsini\ implies periods in the range $-0.09$d $ < \log {\rm P}_{\rm rot} <
0.22$d. By looking at Figure 16 of \citet{strassmeier12}, we see that
there are a few such systems, even with higher rotational velocity.
Again, the high rotational velocity favours a young age, and excludes
the star from being a solar sibling.

To be more consistent, we compared the chemical composition of this star with
the solar chemical composition derived from the SOPHIE spectrum of Ceres. The
abundances derived from the spectrum of the asteroid are different from the
abundances of this star (e.g. there is a difference of 0.19\,dex in Na,
0.24\,dex in Ca, 0.13\,dex in Fe from neutral lines).  These differences in
the chemical composition alone would allow us to exclude this star as a solar
sibling.  The derived age (see Sect. 5) confirms this conclusion.

\subsection{HIP\,79822 ($\eta$\,UMi)} \label{sec_etaumi}

For this bright star which is part of a double system, no chemical
analysis can be found in the literature.  The secondary component is
apparently an M\,4 dwarf \citep[][V=15.3]{Lepine}, for which the determination of
the chemical composition from its spectrum is extremely difficult.
One could constrain the chemical composition of a few elements from
the F-star companion, since these are available.

We would like to point out that our measured radial velocity
is significantly different from most of the radial velocities
reported in the literature for this star, that range from
+35.7\,\kms\ to --20.9\,\kms\ \citep{Hnatek} through -11.8\,\kms\ \citep{Plaskett},
with a more recent measurement of -11.1\,\kms\ provided by CORAVEL
\citep{Birgitta}.
Our radial velocity is compatible with the lowest value found in the
literature and based on an observation taken on June 11 1915
with the Coud\'e spectrograph of the University Observatory
of Vienna \citep{Hnatek}.
Since the star is a binary, it is not
surprising that the radial velocity may be varying.
We encourage further monitoring of the radial velocity of this star.

\citet{casagrande11} state that the star has an effective temperature
of 6820\,K, but our fit of the H$\alpha$ wings suggest a hotter
temperature, \teff\ =6946\,K.  The two \teff\ agree within the
uncertainty of 100\,K that we associate to the determination of the
temperature from the H$\alpha$ wings.  We derived the surface gravity
from the Hipparcos parallax (33.63\,mas, \citealt{vanleeuwen07}),
assuming 1.5\,${\rm M}_\odot$, with an uncertainty of 0.2\,dex.  For
the two effective temperature values, 6820\,K and 6946\,K, we derived
\logg\ of 4.02 and 4.05, respectively.  We fixed the micro-turbulence
at 2\,\kms.

The star has a high \vsini\ of 85\,\kms , in good agreement with other values found in the literature
\citep{Birgitta,Reiners,Schroeder}.
This high \vsini\ does not allow us to analyse this
spectrum using the same techniques we used for the spectra of the
other stars in this project.  It is very difficult to find a single
unblended lines in the spectrum.  To derive the abundances we selected
blends in which one element is the major contributor, or features
where several lines of the same element contribute to the blend.  The
results, in the case the effective temperature is set at 6946\,K, are
reported in Table\,\ref{etaumi}.

When taking \teff\ =6820\,K and \logg\ =4.02 instead, we have very minor
increases in the abundance.  The most affected element is Ba whose
abundance becomes 0.1\,dex higher. The Mg abundance increases by
0.08\,dex; Fe, Cr, Ni abundances result 0.06\,dex higher; and the Ti
abundance is 0.05\,dex higher.  The other elements are less affected.

\begin{table}
\caption{Chemical abundances of HIP\,79822 ($\eta$\,UMi) for \teff\ of 6946\,K.}
\label{etaumi}
\begin{tabular}{lrrrrrr}\hline
\hline
Element &   N    & $A_\odot$ & $A$(X) & [X/H] & $\sigma$ & [X/Fe] \\
\hline
\ion{Mg}{i}  &    3 &  7.54 &  7.44 &--0.10 & 0.13   & +0.00  \\
\ion{Si}{i}  &    3 &  7.52 &  7.26 &--0.26 & 0.22   & +0.16  \\
\ion{Si}{ii} &    1 &  7.52 &  7.63 & +0.11 &        & +0.21  \\
\ion{S}{i}   &    4 &  7.16 &  7.16 & +0.00 &        & +0.10  \\
\ion{Ca}{i}  &    7 &  6.33 &  6.40 & +0.07 & 0.12   & +0.17  \\
\ion{Ti}{i}  &    2 &  4.90 &  4.81 &--0.09 & 0.22   & +0.01  \\
\ion{Ti}{ii} &    1 &  4.90 &  4.79 &--0.11 &        &--0.01  \\
\ion{Cr}{i}  &    1 &  5.64 &  5.45 &--0.19 &        &--0.09  \\
\ion{Fe}{i}  &   53 &  7.52 &  7.42 &--0.10 & 0.14   &  0.00  \\
\ion{Fe}{ii} &    4 &  7.52 &  7.49 &--0.03 & 0.25   &  0.00  \\
\ion{Ni}{i}  &    3 &  6.23 &  6.22 &--0.01 & 0.01   & +0.09  \\
\ion{Ba}{ii} &    2 &  2.17 &  2.26 & +0.09 & 0.39   &  0.16  \\
\hline
\end{tabular}
\end{table}

\section{Uncertainties}

For five of the stars we analysed in this project (HD\,120205,
HD\,136137, HD\,136655, HD\,150202, HD\,171067), we derived the
stellar parameters and the abundances with MyGIsFOS.  \teff\ is
obtained by imposing no slope in [Fe/H] vs. lower energy of the Fe
transitions.  \logg\ is derived from agreement of [Fe/H] from
\ion{Fe}{i} and \ion{Fe}{ii} lines.  For the slope in excitation
energy we can conservatively expect an uncertainty of 50\,K in \teff.
The number of \ion{Fe}{i} lines is large (65-107) and they give a
line-to-line abundance scatter of about 0.1\,dex. \ion{Fe}{ii} lines
are less numerous (between 14 and 16) and they provide a line-to-line
scatter a bit larger than 0.1\,dex (0.11-0.17\,dex).  We can associate
the uncertainty in \logg\ with a disagreement in [Fe/H] from
\ion{Fe}{i} and \ion{Fe}{ii} lines of 0.1\,dex that, for the range in
stellar parameters of these five stars, can be translated into an
uncertainty in \logg\ of 0.2\,dex.  A disagreement of 0.2\,dex from
\ion{Fe}{i} and \ion{Fe}{ii} lines would give a change in \logg\ of
0.3\,dex.

We analysed a spectrum of Ceres to have better constraints on the
uncertainties we can have on the stellar parameters of the five stars
analysed by means of MyGIsFOS.  Comparing the results of Ceres in
Table\,\ref{param} to the stellar parameters of the Sun we can assign
a systematic uncertainty of 80\,K for \teff, 0.1\,dex for \logg,
0.2\,\kms\ for the micro-turbulence $\xi_{micro}$\,and 0.1\,dex for
the metallicity.

Conservatively, we can take the following values as uncertainties in
the stellar parameters for these five stars: 100\,K in \teff, 0.2\,dex
in \logg, 0.2\,\kms\ in micro-turbulence, and 0.1\,dex for the
metallicity.

\section{Stellar activity}

The emission cores of the \ion{Ca}{ii} H\&K resonance lines are routinely used 
as a diagnostic of the chromospheric activity. 
Our spectra are even of sufficient resolution to define the emission line profiles 
itself by, e.g., detecting the central emission-line reversal if present. 
The two minima outside of the emission line, the so-called violet (V) and red (R) footpoints, 
were measured in this paper for both \ion{Ca}{ii} H and K lines. 
We follow the notions and absolute-flux calibration laid out in a previous paper 
by \citet{strassmeier00}. 
It relates high-resolution spectra to the absolute photometric flux in a 5-nm band 
at 395\,nm calibrated solely upon the $B-V$ or $V-R$ color of the star.

We first measure the relative flux of each emission line between the V and R points 
with the help of the splot routine in iraf as well as the 5-nm flux centered at 395\,nm. 
These fluxes are then converted to absolute fluxes by applying the above photometric calibration. 
It is important to remove the photospheric contribution from these fluxes if the pure chromospheric emission is sought. 
Such radiative losses were identified to be the net cooling rate in the chromosphere 
due to these lines \citep{lin:ayr}. 
The expected radiative losses as a function of stellar $V-R$ were taken from that paper 
and subtracted from the measured values. 
The $B-V$ values used in the calibration were obtained from the \teff\ values and 
are later listed in Table\,\ref{evol}. 
Finally, the sum of the corrected absolute emission-line fluxes from H and K is expressed 
in units of the bolometric luminosity $\sigma T_{\rm eff}^4$ 
and yields a value for the total chromospheric radiative loss $R'(HK)$.

Table\,\ref{T-Ca2} summarizes the numerical values. 
We note that for the two targets $\eta$\,UMi and HD\,171067 we did not detect the V/R footpoints in the spectrum. 
Instead, we assumed an emission-line width at the bottom of the resonance line of 0.1\,nm, 
which happened to be identical to the V/R width of HD\,120205. 
The $F'(K)$ and $F'(H)$ values are the corrected absolute chromospheric emission-line fluxes. 
These fluxes are likely not better than 20--30\%\ despite that the relative fluxes 
are determinable to within a per cent from repeated measurements. 
The respective radiative losses $R'(HK)$ are then uncertain by approximately 0.12\,dex.

\begin{table}
\caption{Chromospheric emission properties.}
\label{T-Ca2}
\begin{tabular}{lllll}\hline
\hline\noalign{\smallskip}
Star        & $V-R$ & $F'(K)$ & $F'(H)$ & $\log R'(HK)$ \\
 & (mag) & \multicolumn{2}{c}{(10$^6$ erg\,cm$^{-2}$s$^{-1}$)} &  \dots \\
\noalign{\smallskip}\hline\noalign{\smallskip}
HD\,120205  & 0.715 & 0.40  & 0.30 & --4.81 \\
HD\,136137  & 0.71  & 0.63  & 0.43 & --4.59 \\
HD\,136655  & 0.75  & 0.41  & 0.33 & --4.74 \\
HD\,150202  & 0.71  & 0.85  & 0.71 & --4.42 \\
HD\,171067  & 0.595 & 0.975 & 0.975& --4.42 \\
HD\,147443  & 0.51  & 1.86  & 1.82 & --4.37 \\
$\eta$\,UMi & 0.39  & 2.48  & 2.86 & --4.39 \\
\noalign{\smallskip}\hline
\end{tabular}
\end{table}

\section{Ages, masses, and radii}

Making use of the effective temperatures and metallicities we
determined in this work, the magnitude in band V and the Hipparcos
parallax, we derived the stellar parameters, age, mass, surface
gravity, radius, with the tool {\it PARAM} publicly
available\footnote{http://stev.oapd.inaf.it/cgi-bin/param} presented
by \citet{dasilva06}.  This tool performs a Bayesian estimate, by
comparing the input parameters to the theoretical PARSEC isochrones by
\citet{PARSEC}.  The results are given in Table~\ref{evol}.  The
stars, as expected, are found to be young; except for HIP\,90864 (HD\,171067), they are all younger than the Sun. 
Their respective chromospheric $R'(HK)$ indices in Table~\ref{T-Ca2} 
suggest that these stars are younger than the Sun. 
Individual ages between $\approx$0.5--3~Gyr are suggested when compared to the recent activity-age relation in \citet{pace}.

The mass derived by {\it PARAM} for
HIP\,80124, ${M}=\left( 1.202\pm 0.053\right) {\rm M}_\odot $, is
in perfect agreement with the mass we assumed in
Sect.~\ref{sec_hip80124} to derive the gravity, ${M}=1.2 {\rm
  M}_\odot$.  For $\eta$\,UMi we had assumed ${M}=1.5 {\rm
  M}_\odot$ in Sect.\,\ref{sec_etaumi} to determine the surface
gravity, in very good agreement with the result of {\it PARAM}
${M}=\left( 1.490\pm 0.057\right) {\rm M}_\odot$ of Table\,\ref{evol}.

\begin{table*}
\caption{Stellar parameters.}
\label{evol}
\begin{tabular}{llllll}\hline
\hline
Star                   & Age              & Mass             & \logg            & Radius           & $({\rm B-V})_0$   \\
                       & (Gyr)           & (${\rm M}_\odot$) &  (cgs)           & (${\rm R}_\odot$)  & (mag)               \\
\hline
HIP\,67344 (HD\,120205) & $3.4\pm 3.4$ & $0.856\pm 0.032$ & $4.545\pm 0.027$ & $0.793\pm 0.023$ & $0.884\pm 0.021$ \\ 
HIP\,74858 (HD\,136137) & $1.1\pm 0.2$ & $2.111\pm 0.151$ & $2.812\pm 0.053$ & $9.151\pm 0.603$ & $0.938\pm 0.021$ \\ 
HIP\,75132 (HD\,136655) & $4.0\pm 3.7$ & $0.869\pm 0.032$ & $4.534\pm 0.030$ & $0.808\pm 0.028$ & $0.926\pm 0.025$ \\ 
HIP\,81284 (HD\,150202) & $0.9\pm 0.3$ & $2.231\pm 0.223$ & $2.772\pm 0.091$ & $9.848\pm 1.351$ & $0.938\pm 0.024$ \\ 
HIP\,90864 (HD\,171067) & $7.0\pm 4.1$ & $0.901\pm 0.039$ & $4.436\pm 0.035$ & $0.922\pm 0.030$ & $0.758\pm 0.019$ \\ 
HIP\,80124 (HD\,147443) & $1.8\pm 1.4$ & $1.202\pm 0.053$ & $4.292\pm 0.063$ & $1.257\pm 0.096$ & $0.591\pm 0.022$ \\ 
HIP\,79822 ($\eta$\,UMi)& $1.6\pm 0.3$ & $1.490\pm 0.057$ & $4.059\pm 0.036$ & $1.830\pm 0.061$ & $0.418\pm 0.022$ \\ 
\hline
\end{tabular}
\end{table*}

\begin{figure}
\includegraphics*[width=77mm]{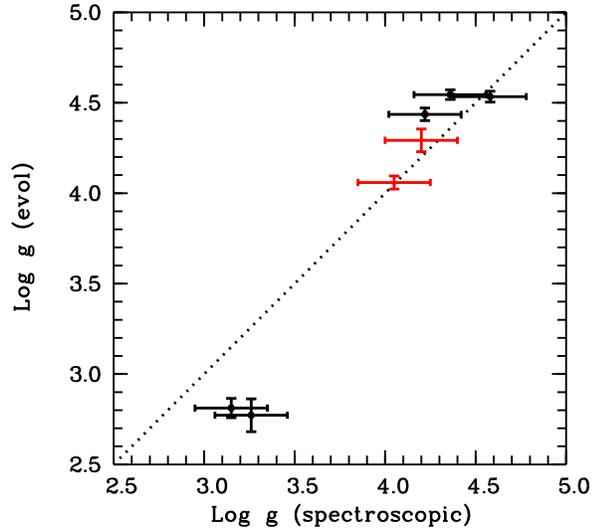}
\caption{The spectroscopic surface gravity we derived from the chemical analysis
of the SOPHIE spectra versus the gravity estimated from parallaxes.}
\label{logg2}
\end{figure}

For two of the stars in our analysis (HIP\,80124 and HIP\,79822) the
surface gravity was derived from the parallax, although the ionization
equilibrium of \ion{Fe}{i} and \ion{Fe}{ii} we obtained was largely
within the uncertainties. Although, in Fig.\,\ref{logg2}, the comparison of the
spectroscopic surface gravity derived in our analysis to the evolution
surface gravities is fair only for the other five stars (black symbols in the figure).
For the two stars (red symbols, HIP\,80124 and HIP\,79822) the two
\logg\ values are both based on the parallax, and as expected,
the agreement is very good. Also for the other three
unevolved stars the agreement is within the uncertainties.  The
agreement is worse for the two sub-giant stars (see
Fig.\,\ref{logg2}).

\section{Conclusions}

We analyzed a sample of seven young, solar metallicity stars for which
the chemical analysis was lacking or not complete.  We compared our
results with the stellar parameters available in the literature,
finding a general agreement.  For five stars we derived the stellar
parameters and the detailed chemical composition with our automatic
code MyGIsFOS.  For the two stars with high rotational velocities
(HIP\,80124 and HIP\,79822) we derived the stellar parameters by other
methods: \teff\ from the wings of H$\alpha$ and gravity from parallax
and assumptions about the stellar mass. The chemical composition of
these two stars were derived from EW measurements or line profile
fitting of the available features.

Five of the stars show emission in the core of the \ion{Ca}{ii}-H and
-K lines. Their chromospheric emission-line fluxes are in agreement with ages between 0.5--3~Gyrs. 
For HIP\,74858 and HIP\,79822 no previous chemical analysis
had been done.  For the latter star, a chemical analysis is
particularly interesting because this star has a cool companion dwarf
of type M\,4.  Deriving the chemical composition for stars of this
stellar type is very challenging and the chemical composition of the
F-type companion would allow to improve our understanding of spectra
of these cool stars.

In the sample of stars presented, two of them showed a strong
\ion{Li}{i} doublet from which we could derive the Li abundance. This
was done using detailed 3D model atmospheres and by considering
departures from LTE.  The chemical pattern of HIP\,80124 is not
consistent with the solar composition derived from the spectrum of
Ceres, which excludes it from being a solar sibling, as suggested by
\citet{brown10}.  Its age is also found to be much younger than that
of the Sun.

\acknowledgements
We are grateful to F. Bouchy for redetermining off-line the radial
velocity of HD\,136655 from our SOPHIE spectrum.
The project was funded by FONDATION MERAC.

\newpage

\appendix

\section{The 3D+NLTE abundance correction}
\label{sec:A}
We derive the ``3D+NLTE abundance correction'' (hereafter $\Delta_{\rm 3DNLTE}$)
from an independent fitting procedure involving only synthetic spectra:
for a given 3D hydrodynamical stellar atmosphere (taken from the CIFIST
grid), and a given lithium abundance, A$^\ast$(Li), we compute a synthetic
spectrum of the wavelength region around the \ion{Li}{i} line at 670.8\,nm,
taking into account the 3D non-LTE effects for lithium, while treating the
blending lines in LTE. This 3D non-LTE synthetic spectrum is then fitted
by a grid of 1D LTE LHD reference spectra in exactly the same way as the
observed spectrum is fitted with a grid of 1D LTE ATLAS9 synthetic spectra.
The 1D LTE lithium abundance obtained from the best fit, A(Li)$_{\rm 1DLHD-LTE}$,
is then used to define the 3D+NLTE abundance correction as
\begin{equation}
\Delta_{\rm 3DNLTE}=\mathrm{A}^\ast(\mathrm{Li})-\mathrm{A(Li)}_{\rm 1DLHD-LTE}\, .
\end{equation}

The 1D reference model used for this procedure is a
hydrostatic plane-parallel model atmosphere computed
with the LHD code (see \citealt{CaffauLudwig2007})
for exactly the same stellar parameters as the corresponding
3D model. It employs the same micro-physics and numerics (opacities,
equation-of-state, radiative transfer scheme) as the \cobold\ models.
The main advantage of using this kind of 1D reference models is that they
are differentially comparable to the associated 3D model, the only difference
being the treatment of the convective energy transport: mixing-length theory
in the LHD models (here assuming a mixing-length parameter equal to 1.00)
versus 3D hydrodynamics in \cobold.

The 3D+NLTE correction defined in this way takes into account the combined
effect of 3D convection (stellar granulation) and departures from LTE (in the
lithium line formation) on the spectroscopic abundance determination.  In
principle, this 3D+NLTE correction depends on
the stellar parameters and on the lithium abundance A$^\ast$(Li).  However, we
point out that $\Delta_{\rm 3DNLTE}$ is a smooth and slowly varying function
of these parameters. It is therefore sufficient to know this function at a few
points covering the region of interest in the Hertzsprung-Russell diagram
and the relevant range of Li
abundances. The correction can then be obtained by interpolation for any
particular point in the parameter space.

For a particular object, the 3D+NLTE correction is evaluated for the
same stellar parameters as the ATLAS9 model used for fitting the observed
spectrum, and for an input lithium abundance A$^\ast$(Li) defined by the
implicit relation

\begin{equation}
\Delta_{\rm 3DNLTE}(\mathrm{A}^\ast(\mathrm{Li})) =
\mathrm{A}^\ast(\mathrm{Li}) - \rm A(Li)_{\rm ATLAS-LTE}\, .
\end{equation}
Once the 3D+NLTE correction is known, the final lithium abundance is
obtained from:
\begin{eqnarray}
{\rm A(Li)_{\rm 3DNLTE}} &=& \rm A(Li)_{\rm ATLAS-LTE}+\Delta_{\rm
  3DNLTE}(A^\ast(\mathrm{Li})) \nonumber \\
                      &=& \mathrm{A}^\ast(\mathrm{Li}) \, .
\end{eqnarray}

\end{document}